\documentclass{article}

\usepackage{arxiv_settings}

\usepackage[utf8]{inputenc} % allow utf-8 input
\usepackage[T1]{fontenc}    % use 8-bit T1 fonts
\usepackage{hyperref}       % hyperlinks
\usepackage{url}            % simple URL typesetting
\usepackage{booktabs}       % professional-quality tables
\usepackage{amsfonts}       % blackboard math symbols
\usepackage{nicefrac}       % compact symbols for 1/2, etc.
\usepackage{microtype}      % microtypography
\usepackage{lipsum}
\usepackage{fancyhdr}       % header
\usepackage{graphicx}       % graphics
\usepackage{natbib}
\graphicspath{{Images/}}     % organize your images and other figures under media/ folder

\usepackage{algorithm}
\usepackage{amssymb}
\usepackage{mathrsfs}
\usepackage{amsthm}
\usepackage{amsmath}
\usepackage{multirow}
\usepackage{booktabs}
\usepackage{algpseudocode}
\usepackage{xcolor}

\DeclareMathOperator{\argmin}{argmin}

\title{Privacy Re-identification Attacks on Tabular GANs}

\author{
  Abdallah Alshantti\textsuperscript{*}, $\,$ Adil Rasheed, $\,$ Frank Westad \\
  Department of Engineering Cybernetics \\
  Norwegian University of Science and Technology \\
  Trondheim 7034, Norway \\
  \textsuperscript{*} \texttt{abdallah.a.s.alshantti@ntnu.no} \\
}

\begin{document}
\maketitle

\begin{abstract}
%The generation of synthetic data is increasingly contributing to the acceleration of knowledge and information sharing with domain experts and the public. The use of synthesised data is particularly desired for representing the distribution and the semantics of confidential data while preserving the privacy of data records. However, recent studies demonstrated that generative models are subject to overfitting and thus may potentially leak sensitive information from the training data. In this work,
Generative models are subject to overfitting and thus may potentially leak sensitive information from the training data. In this work. we investigate the privacy risks that can potentially arise from the use of generative adversarial networks (GANs) for creating tabular synthetic datasets. For the purpose, we analyse the effects of re-identification attacks on synthetic data, i.e., attacks which aim at selecting samples that are predicted to correspond to memorised training samples based on their proximity to the nearest synthetic records. We thus consider multiple settings where different attackers might have different access levels or knowledge of the generative model and predictive, and assess which information is potentially most useful for launching more successful re-identification attacks. In doing so we also consider the situation for which re-identification attacks are formulated as reconstruction attacks, i.e., the situation where an attacker uses evolutionary multi-objective optimisation for perturbing synthetic samples closer to the training space. The results indicate that attackers can indeed pose major privacy risks by selecting synthetic samples that are likely representative of memorised training samples. In addition, we notice that privacy threats considerably increase when the attacker either has knowledge or has black-box access to the generative models. We also find that reconstruction attacks through multi-objective optimisation even increase the risk of identifying confidential samples.
\end{abstract}

% keywords can be removed

\section{Introduction}

% \begin{itemize}
%     \item Data and its value - open-source, collaborations, need for data
%     \item synthetic data and methods to generate syn data - bayesian models, vae, gans (powerful)
%     \item Privacy concerns of GANs and motivation for this study - outlining the emphasis of privacy attacks on GANs
%     \item highlighting that the attacks are crafted from the perspective of an adversary and such attacks are quantified
%     \item defence mechanisms to protect against such attacks
% \end{itemize}

%Data sharing internally within members of an organisation, or externally between various institutions is increasingly prevalent and contributes to facilitating the understanding of domain knowledge. Accessible and reproducible research and open source codes are also becoming more relevant for facilitating knowledge within the research community and speeding up the pace at which certain sectors are developing at. Furthermore, private organisations are more actively engaged in reproducible research as it demonstrates their commitment and selflessness towards solutions that extend the benefits beyond the organisations themselves.

In recent years, plenty of effort has been dedicated to generating synthetic data as a means of facilitating knowledge exchange whilst maintaining some disparity to protect confidential information. In essence, synthetic data is designed to capture the properties and the general structure of the original data, while simultaneously obscuring the sensitive attributes in the data. Historically, synthetic data was created by omitting or anonymising identifier features in a dataset. Alternatively, modifying sensitive features by adding noise was also considered as an alternative to anonymisation. However, both such approaches have been proven insufficient as sensitive information can be still recovered through de-anonymisation~\citep{narayanan2008robust} and de-noising~\citep{agrawal2000privacy} techniques. More recently, generative models based on neural networks have become widespread techniques for data synthesis, generating new records by estimating the distribution of a given dataset. In particular, generative adversarial networks (GANs)~\citep{goodfellow2014generative} are desired primarily due to their superior performance in approximating the distribution of the data and producing highly realistic data records.  

Tabular data is prevalent in various domains such as the healthcare, finance, e-commerce and cybersecurity fields. Mixed-type data entails that the data consists of binary, continuous and discrete features, which are often challenging to represent by machine learning and deep learning models \citep{popov2019neural}. While early tabular GANs struggled with handling categorical attributes \citep{choi2017generating, park2018data}, more recent tabular GANs have exhibited more success in representing and reproducing categorical features \citep{zhao2021ctab, alshantti2024castgan}. In addition, GANs demonstrated unprecedented capabilities in capturing the distribution of data features and the correlations between them.

Concerningly though, private tabular data often encapsulates sensitive information about individuals or records. To train models that overfit on the data induces then a privacy risk, since such overfitting may be due to some form of memorising data samples by the models. Releasing such models, or even simply enabling users to query random samples on such models can thus give rise to leaking partial or full confidential information about training data records. The risks of overfitting by discriminative models where the primary prediction tasks are classification or regressions have been well highlighted and considered by \cite{song2021systematic}. On the other hand, privacy risks associated with overfitting generative models have been much less studied. This is supported by the fact that detecting overfitting on GANs is far from being a straightforward task, as confidence values about overfitting cannot be directly obtained \citep{hayes2019logan, chen2020gan}. One method for analysing the susceptibility of generative models to privacy risks is through the implementation of privacy attacks.

Privacy attacks have been extensively studied in the context of discriminative models. In particular, membership inference attacks attempt to infer whether a given record in a holdout dataset was used for training a discriminative model \citep{shokri2017membership}. Since then, membership inference attacks have been explored within different contexts. Another type of privacy attacks is the model-inversion attack \citep{fredrikson2014privacy}, where an adversary attempts to use the model's output to recover training datapoints. While most of the pertinent research on privacy attacks remains to be aimed at discriminative models, privacy attacks have been eventually extended to the generative domain.
%Similar to inference attacks, the concept of model inversion attacks was further studied to consider the impact of different available resources or information \citep{hidano2017model, }.

Membership inference attacks against GANs were implemented by \cite{hayes2019logan}, where confidence values about a given record were obtained from the discriminator of the GANs. Meanwhile, \cite{hilprecht2019monte} formulated two types of membership attacks and included variational autoencoders (VAEs) as a target generative model. \cite{chen2020gan} conducted membership inference attacks on multiple generative models and datasets while demonstrating the factors that influence the success of inference attacks. Inspired by these existing works, we posit that there is a need for considering the implications of potential privacy risks on tabular data. We therefore hypothesise that the increased accessibility of tabular GAN models can threaten the privacy of sensitive information. Moreover, from intuitive perspectives the risk seems heightened for smaller and lower-dimensional datasets, and for mixed-type datasets where categorical features can take a finite range of values. 

% Need to rewrite the motivation part below

Finally, to the best of our knowledge the following question remains unanswered: \emph{given access to a synthetic dataset, can an attacker exploit the data for predicting which synthetic samples are likely to be identical to or leak a substantial amount of information from corresponding training samples?} A positive answer to such a question would imply that privacy attacks may be successful even when an attacker does not have a holdout set for conducting membership inference attacks on. In this case, the growing demand and accessibility of generative models, together with the availability of synthetic data and generative models may enable malicious actors to perform successful attacks without necessarily having access to a query set for inference. Given this risk, we foresee the need for studying the possibility of re-identification attacks on synthetic datasets for recovering training datapoints which were used for training generative models. In other words, we consider the previous question as the core research one for this work, and offer the following contributions:

\begin{itemize}
    \item We describe multiple attacking scenarios based on the potential access levels for the attackers. Namely, in addition to the possession of synthetic samples, an attacker might also have knowledge of the generative model's architecture, have black-box access to the trained model's API and potentially a machine learning prediction model trained on either the synthetic or the private data. 
    \item We conduct re-identification attacks using the synthetic samples in the access settings mentioned above. In this context, the re-identification attacks can be referred to as selection attacks, where the attacker selects the most densely surrounded synthetic datapoints as the candidate samples that were possibly memorised by a generative model.
    \item We further analyse the effects of formulating re-identification attacks as reconstructions attacks, i.e., where an attacker uses for its purposes evolutionary multi-objective optimisation to perturb the candidate synthetic samples. In this case, the attacker attempts to reduce the proximity of a synthetic sample to its neighbouring synthetic samples and also reduce the prediction error of its target class. 
    \item We quantitatively evaluate the privacy risk of re-identification attacks as selection or reconstruction ones. In addition, we adapt the existing membership inference attacks from literature as re-identification ones, and compare the success rate of our attacks against the baselines.
\end{itemize}

The remainder of the paper is structured as follows. Section 2 outlines the fundamentals for GANs, privacy attacks and privacy defenses. We formulate and describe the privacy re-identification attacks in Section 3. In Section 4, we outline our experimental setup including the datasets used, generative models considered and our evaluation criteria, while our results are presented in Section 5. We finally provide a comprehensive summary of the related works in Section 6 and we conclude this work in Section 7.

\section{Background}

%Potential of deep learning, threats against deep learning, generative models

\subsection{GANs}

Among generative models, GANs \citep{goodfellow2014generative} have in recent times become the most widely used approach for creating synthetic data. Ultimately, this is attributed to superiority of GANs in approximating the probability distribution, in contrast to statistical generative models such as Bayesian models \citep{koller2009probabilistic}, Hidden Markov models \citep{rabiner1989tutorial}, Gibbs sampling models \citep{park2014pegs} and other deep generative models such as denoising autoencoders \citep{gondara2018mida} and variational autoencoders \citep{kingma2014semi}. As such, GANs have been widely adopted in various domains such as image generation \citep{zhu2017unpaired, karras2019style}, time-series generation \citep{esteban2017real, yoon2019time} and tabular data generation \citep{choi2017generating, xu2019modeling, engelmann2021conditional}.

The classical GAN is composed of two neural networks competing against each other in an adversarial setting. Namely, a generator $G$ takes a random noise vector $z$ as an input, and produces fake data samples as the output. The other neural network is a discriminator $D$ that receives the real data in addition to the fake data from the generator as inputs, and attempts to distinguish the real samples from the synthetic samples. As the two components aim to maximise their gain in a min-max game, the generator becomes increasingly skilled during the training process at producing samples that closely resemble the real data based on the feedback it receives from the generator, meanwhile, the discriminator improves its capability in discriminating between both inputs. The training process can be resembled as:

\[
    \min_{G} \max_{D} V(D,G) = \mathbb{E}_{x \sim p_{data}(\boldsymbol{x})}[log(D(\boldsymbol{x})] + \mathbb{E}_{z \sim p_{z}(\boldsymbol{z})}[log(1-D(G(\boldsymbol{z})))
\]

where $p_{data}$ is the distribution of the real data and $p_{z}$ is the distribution of the noise sample.

In addition to generating realistic output without explicitly sampling a parametric likelihood function on the data feature space, GANs also eliminate the one-to-one relationship between the synthetic data and the original data, thus reducing the likelihood of sensitive attribute leakage \citep{park2018data}. Nevertheless, it has been demonstrated that GANs are still widely susceptible to privacy attacks \citep{webster2019detecting}. To this end, privacy attacks which were originally devised for deep learning discriminative models can also be adapted for targeting GANs. Such attacks include membership inference attacks \citep{shokri2017membership}, co-membership inference attacks \citep{liu2019performing}, training class inference attacks \citep{yang2019neural}, property inference attacks \citep{rigaki2020survey} and model inversion attacks \citep{fredrikson2014privacy}.

% More recently, the threat against GANs have been studied. (few citations, only highlighting the threats and motivation).

%Generative models are explored in ... 

\subsection{Membership Inference Attacks}

Membership inference attacks (MIAs) were first devised by \cite{shokri2017membership}, in which classification models are targeted in a black-box setting. In MIA, an attacker is provided with a query dataset from an unknown source and attempts to identify the data records that were used for training a machine learning model. Whereas, in white-box membership inference attacks the perpetrator has access to the internals of the training model and uses this knowledge to make better-informed decisions about the membership of the records in the query set. It has been demonstrated that white-box MIAs on a neural network's stochastic gradient descent optimizer are far more powerful than the standard black-box attacks \citep{nasr2019comprehensive}. Moreover, membership inference is increasingly explored in the federated learning domain, in which a model is trained in a decentralised manner by several actors \citep{melis2019exploiting}. Federated learning can however introduce data leakages which add up to the privacy concerns. In addition, it has been observed that while overfitting does contribute to the data leakage \citep{shokri2017membership}, it was also shown that a well-generalisable model is still largely susceptible to effective membership attacks \citep{yeom2017unintended,long2018understanding}, thus motivating for further studies on membership inference attacks. 

More recently, the application of membership inference attacks has also been extended to generative models. \cite{hayes2019logan} presented the first study of membership inference attacks on GANs whereby it was found that white-box attacks can exploit the overfitting in generative models, thus shedding light on the magnitude of privacy leakage issue in generative applications. The membership inference attacks are further bolstered by \cite{hilprecht2019monte}, who formulated a new type of MIAs based on Monte Carlo and demonstrated their successfulness against GAN models. Meanwhile, \cite{chen2020gan} comprehensively studied MIAs on various GAN implementations and demonstrated that full white-box MIAs are persistently more effective than grey-box and black-box attacks.

\subsection{Model Inversion Attacks}

In model inversion attacks, an adversary, given a machine learning model, aims to retrieve the original input used for training the model. The study of model inversion attacks traditionally assumes that an attacker relies on white-box attacks to extract the model parameters and uses them to revert the functionality of the victim model for exposing the raw data \citep{fredrikson2014privacy}. Meanwhile, \cite{hidano2017model} proposed model inversion attacks without the knowledge of non-sensitive attributes which achieve comparable performance to the attacks in \citep{fredrikson2014privacy}. Similarly, \cite{tramer2016stealing} proposed model extraction attacks in which an attacker launches attacks in a black-box setting to emulate the functionality of a discriminative model without relying on prediction confidence values. Meanwhile, \cite{yeom2017unintended} studied the relationship between model inversion attacks and membership inference attacks, and formulated a new type of attack that can threaten generalisable models. Synonymous to model inversion attacks, \cite{cai2021generative} defines reconstruction attacks as those that attempt to recover the raw training data given the model and additional auxiliary information.

% Additionally, \cite{fredrikson2014privacy} studied the link between applying model inversion attacks and using differential privacy as a defence mechanism, where the authors found that differential privacy with specific budget parameters can lead to more resilient models against model inversion threats.

In contrast, model inversion attacks have only been explored to a limited extent in the context of generative models. \cite{zhang2020secret} developed a framework that utilises a generative adversarial network coupled with some auxiliary knowledge for launching inversion attacks on deep neural networks used for image classification. Whereas, CPGAN is an approach proposed by \cite{tseng2020compressive} as a privacy preserving pre-processing step for compressing representations of image datasets prior to training classification models. Conversely, \cite{aivodji2019gamin} proposed using generative adversarial networks for creating black-box inversion attacks against a victim convolutional neural network classification model.

\subsection{Defences}

% Song and Mittal -> MemGuard
% Logan (hayes et al.) ->
% TableGAN -> k-anonymity
% Adversarial Regularisation - Nasr et. al

Mitigating against privacy attacks for discriminative models and generative models has been explored in the literature. \cite{li2006t} proposed the t-closeness concept as a privacy-preserving technique by ensuring that the distribution of a sensitive feature in a categorical group is similar to the distribution of the feature in the entire dataset. Weight normalisation \citep{salimans2016weight} and dropout \citep{srivastava2014dropout} were considered by \cite{hayes2019logan} as regularisation mechanisms for preventing overfitting and subsequently hampering the impact of membership inference attacks. However, it has been observed that both techniques significantly slow down the training process and can contribute to training instability. \cite{nasr2018machine} introduced membership inference adversarial training as part of the standard target classifier training to induce regularisation that protects against membership inference attacks. Meanwhile, MemGuard was proposed by \cite{jia2019memguard} where the predictions of a target model are obscured with carefully crafted noises to reduce the effectiveness of membership inference attacks while minimally impacted the classification predictions. 

Differential privacy is a concept that has been proposed by \cite{dwork2008differential}, which entails that any two datasets differing by a single observation are considered adjacent. Thus an algorithm is considered $( \epsilon,  \delta )$ deferentially private if it meets the adjacency condition, where $\epsilon$ represents the privacy budget parameter and $\delta$ is a term that quantifies the violation of differential privacy. The application of differential privacy in deep learning has been explored extensively by \cite{shokri2015privacy} and \cite{abadi2016deep}, where it was found that differential privacy is capable of providing sufficient privacy guarantees in most cases. Nevertheless, it has been demonstrated that differential privacy exhibits a significant trade-off between model accuracy and privacy mitigation \citep{ shokri2015privacy, jayaraman2019evaluating}.

% \subsection{Privacy in generative models}

% diff priv, MIAs, canonical .... PATEGAN, DPGAN, PrivBN, ...

\section{Methodology}

In this section we describe the attack types that are instigated by an adversary, and how does this contribute to the bigger picture. The main research question is therefore: \emph{Given the synthetic data, to what extent can the attacker use this synthetic data to re-identify the original training samples?}

\subsection{Problem Definition}

Let $\mathcal{D} = \lbrace \mathbf{x}, y \rbrace$ be a private dataset, where $\mathbf{x} = \lbrace \mathrm{x}_1, \dotsc, \mathrm{x}_m \rbrace$ are the $m$ predictive features of the dataset, and $y \in \lbrace 1, \dotsc , C \rbrace$ is the target attribute for a classification task of $C$ classes or $ y \in \mathbb{R}^{1}$ in the case of a regression task. Within a GAN model, synthetic samples are produced by the generator $G$, in which the generator typically takes a noise vector $z \sim \mathcal{N} ( 0,1 )$ as input. The synthetic dataset can therefore be denoted as $ \mathcal{D}' = G(z) $. Similar to the structure of the private dataset, the synthetic data can be represented as $\mathcal{D}' = \lbrace \mathbf{x}', y' \rbrace$. Meanwhile, a machine learning model for predicting the target feature can be denoted by $\mathcal{M}(\mathbf{x}) \mapsto \hat{y}$ if the model is trained on the real private data or $\mathcal{M}(\mathbf{x}') \mapsto \hat{y'}$ if the model is trained on the synthetic data.

The idea behind re-identification attacks is to find synthetic samples that are in very close proximity to other synthetic datapoints. A generative model that overfits the training data tends to memorise specific real datapoints \citep{ganev2022robin}, hence, generating multiple instances of synthetic samples that can be almost identical to a memorised sample. To find the closely packed synthetic samples, an adversary uses the $k$-nearest neighbours technique to find the closest neighbours and their distances from each query sample in the synthetic dataset. For a given synthetic sample $d'_0 \in \mathcal{D}'$, we define its $k$-nearest neighbours as:

\begin{equation}
    q_{d'_0} = \lbrace d'_1, \dotsc, d'_k \rbrace
\end{equation}

in which the $k$-nearest-neighbours can be determined by:

\begin{equation}
    \lbrace d'_1, \dotsc, d'_k \rbrace = \underset{d' \in \mathcal{D'} \setminus d'_0}{\argmin_{1, \dotsc, k}} \; \lVert d'_0 - d' \rVert^{2}_{2} 
\end{equation}

In addition, the vector of the $k$-nearest-neighbours distances $r_{d'_0}$ can be denoted by:

\begin{equation}
    r_{d'_0} =  \lbrace \lVert d'_0 - j \rVert^{2}_{2}  \rbrace \;\; \text{for} \;\; j \in q_{r'_0}
\end{equation}

and subsequently the harmonic mean of distances of the nearest $k$-nearest-neighbours to the query synthetic sample is represented by:

\begin{equation}
    \bar{r}_{d'_0} = \frac{\lvert r_{d'_0} \rvert}{\sum\limits_{r \in r_{d'_0}} \frac{1}{r}}
\end{equation}

For targeting the most densely packed synthetic samples, the attacker ranks the synthetic samples in $\mathcal{D}'$ in increasing order with respect to their sum of distances to their neighbours $\bar{r}_{d'_0}$. In our study, we posit that the attacker aims to recover a subset of the original training set rather than whole training set as they aim to direct their focus towards training datapoints that were memorised by the generative model. While it is reasonable to assume that a subset accounting for 1\% to 10\% of the training set size is sufficient, we set the subset size to account for 5\% of the training set size, $\tau = 0.05$, as this ratio produces an adequate number of recovered samples. Furthermore, while determining the optimal number of nearest neighbours $k$ is a non-trivial task, we argue that adversaries carrying out the attacks in a black-box setting are unable to perform fine hyperparameter tuning since they do not possess the means for evaluating the performance of their reconstruction attacks. Therefore, the number of nearest neighbours $k$ is set to 5, which represents a satisfactory trade-off between the size of the various training sets and the precision of adversary attacks.  

During the process of selecting the top 5\% synthetic samples as the recovered samples, the attacker omits the synthetic samples that appeared as neighbours to previously selected samples. This is done to diversify the training samples attacked and to avoid the repetitive selection of synthetic samples that could potentially correspond to the same private datapoint. The reconstruction attack process is demonstrated in Algorithm \ref{alg:attack_selection}. 

\begin{algorithm}
\caption{Recovered Samples Selection}
\label{alg:attack_selection}
\begin{algorithmic}
\Require $N_{train} \,$: size of training set \newline
     \hspace*{2em} $\mathcal{D'} \,$: synthetic dataset \newline
     \hspace*{2em} $\tau = 0.05 \,$: ratio of reconstructed samples \newline
     \hspace*{2em} $k = 5 \,$: number of nearest neighbours \newline
     \hspace*{2em} $\bar{R} = \{\} \,$: harmonic mean set \newline
     \hspace*{2em} $Q = \{\} \,$: nearest neighbours set \newline
     \hspace*{2em} $\mathcal{R} = \{\} \,$: reconstructed samples set \newline
     \hspace*{2em} $V = \{\} \,$: discarded samples set
% \Require query sample $d'_0 \,$, \newline 
%     \hspace*{2em} objectives $F=\{f_1, f_2\} \,$, \newline 
%     \hspace*{2em} constraints $G=\{g_1, \cdots, g_{m_{cat}} \}\,$,  \newline
%     \hspace*{2em} number of generations $N_{gen} \,$, \newline
%     \hspace*{2em} population size $L \,$, \newline
%     \hspace*{2em} MCDM weight vector $\Omega = \{ \omega_1, \omega_2 \}$
\Ensure $\mathcal{R}$, reconstructed samples set 
\State $N_{recon} \gets N_{train} \times \tau$
\State $knn\_model_{\text{Train}}(\mathcal{D'}, k) $
\For{$d' \in \mathcal{D'}$}
\State $q_{d'}, r_{d'} \gets knn\_model_{\text{Obtain}}(d')$
\State $\bar{r}_{d'} \gets harmonic\_mean(r_{d'})$
\State $Q \gets Q \cup q_{d'}$
\State $\bar{R} \gets \bar{R} \cup \bar{r}_{d'}$
\EndFor
\State $\mathcal{D'}, Q, \bar{R} \gets sort\_on(\bar{R})$
\For{$d' \in \mathcal{D'}$}
\If{$d' \notin V$}
\State $V \gets V \cup q_{d'}$
\State $\mathcal{R} \gets \mathcal{R} \cup d'$
\If{$\lvert \mathcal{R} \rvert == N_{recon} $}
\State terminate
\EndIf
\EndIf
\EndFor
\end{algorithmic}
\end{algorithm}

\subsection{Attacks on GAN model}

In principle, it is possible for attackers to have different privileges and access levels that could present more serious privacy risks to the data used for generative model training. The prevalence of open source computing, accessibility of model APIs and the availability of partial auxiliary information entail that attackers can exploit such privileges for malicious intent. Subsequently, it is of interest to analyse how much benefit the attackers gain with the possession of additional resources. We therefore consider three scenarios where an attacker might launch reconstruction attacks on the synthetic model.

\textbf{Level I Attack}: In this scenario, an attacker only has access to a synthetic dataset produced by a GAN model.

\textbf{Level II Attack}: This is an attack typically committed by an individual with access to the untrained generative model (e.g. via open-access repository or knowledge of the model's architecture from a research article) and access to the synthetic data. The attacker uses the synthetic data for training the same model and is capable of generating an indefinite number of samples. Then, the attacker attempts to find the new synthetic samples that are close to the training space via black-box selection attacks.

\textbf{Level III Attack}: The attacker has access to the trained generative model, and can use it to generate an indefinite number of samples. This attack is instigated by someone with access to the generative model's API e.g. someone with basic authorisation or someone working at an external organisation. As with Level II attacks, by generating a large number of samples the attacker attempts to find the synthetic samples that are in close vicinity to one another and selects them as the re-identified samples. However, in this case, the attacker has the advantage of generating a large number of samples from the generative model trained on the private dataset. 

We highlight that in all the aforementioned scenarios, the size of the original synthetic data is equivalent to the number of training samples and the attackers know the size of the training set. While Level II and Level III can fundamentally be used for generating an infinite number of samples, we set the size of synthetic datasets generated by both attacks to be $10 \times \lvert \mathcal{D} \rvert$, which is sufficiently large for reconstructing attacks without incurring extremely high computational cost. 

% \textbf{Level III attack}: In this scenario, the attacker has access to the synthetic data and unrestricted access to the generative model. The attacker uses this to query a synthetic sample into the discriminator, and by obtaining the discriminator's score for that datapoint, the attacker recursively modifies the features of the datapoint until the discriminator outputs it is highest score (datapoint close to real space). This attack is typically committed by an internal individual with high privileges or possible developers who are attempting to recover deleted training data based on the synthetic data and trained model. 

% \subsection{Auxiliary Models}

% We also consider the case where an attacker might have access to a predictive model for the dataset used in the study for their advantage:

% \textbf{None}: The attacker has no access to predictive models.

% \textbf{Syn}: The attacker has access to a predictive model fitted on the synthetic data. 

% \textbf{True}: The attacker has access to a predictive model fitted on the real data

\subsection{Research Questions}

Based on the objectives of this study, the following research questions are presented:

\textbf{RQ1}: \emph{To what extent does the privacy risk increase when attackers possess higher access levels to the generative models?}

The aforementioned research question relates to quantifying the success an attacker has if they are able to generate synthetic samples without restriction, in comparison to recovering private samples with access to only a synthetic dataset. In more detail, the main aim of the question is to investigate whether Level II and Level III attacks are generally more effective than a Level I attack, in which an attacker only works with the original synthetic data they have. In addition, we consider analysing the impact of having access to the trained generative model for generating a large number of samples, in contrast to training a tabular generative model on the original synthetic data. 

\textbf{RQ2}: \emph{Does utilising a discriminative machine learning model help in launching more effective attacks?}

It is also interesting to assess whether access to a machine learning model for predicting the target variable can provide an additional advantage for the attacker in reconstructing the private data instances. In particular, a classification or a regression model that performs well in predicting the target variable could indicate more certainty of a given synthetic sample being closely attributed to the original data space. Meanwhile, a synthetic sample that is not representative of any of the training examples is likely to be poorly predicted by the machine learning model. For utilising a prediction model in reconstruction attacks, the adversary assigns weights to the prediction error and the harmonic mean to the nearest neighbours. 

\textbf{RQ3}: \emph{How effective is the use of evolutionary multi-objective optimisation in reconstructing more precise attacks?}

In addition to devising re-identification attacks based on the vicinity of nearby synthetic samples, it is intriguing to investigate whether further perturbing a synthetic sample can direct an attacker closer towards a private training datapoint. By setting the distance to neighbouring synthetic samples and the predicted value of the machine learning model as the two objectives of the multi-objective optimisation problem, the evolutionary algorithm perturbs the synthetic feature space to optimise the two provided objectives. Therefore, we can infer the impact of evolutionary algorithms in the reconstruction process by quantitatively analysing the reconstructed samples at the end of the optimisation process.

\subsection{Evolutionary Optimisation}

By using evolutionary algorithms, a single-objective optimisation problem is solved using a population-based approach in which an initial population is selected and is continuously evolved to minimise the objectives while adhering to the problem's constraints, if present. The final valid solution is a single point in space which both minimises the defined objective and adheres to the problem's defined constraints.

Meanwhile, in multi-objective optimization, a number of defined objectives are optimised simultaneously. The objectives can be conflicting and competing in nature, and this gives rise to an increased complexity in solving the optimisation problem. As opposed to single-objective optimisation where the solution is essentially a single point in space, the final optimal solutions of multi-objective solutions are called Pareto-optimal solutions, which are trade-off solutions to the objectives to be minimized. Subsequently, while a non-trivial task, a single solution can be chosen from the Pareto-optimal solutions based on the desired trade-off between the objectives.

\subsubsection{Evolutionary Algorithm}

A vast number of evolutionary algorithms exist in the literature, and several comparative studies were conducted for determining the suitability of evolutionary algorithms in solving real-world problems \citep{zitzler1999multiobjective, zhang2007moea, ishibuchi2016performance}. To this end, the choice of an evolutionary algorithm for a specific application largely depends on the nature of the problem, the desired number of tuning parameters and the domain knowledge of the objectives and trade-offs between them. 

In this work, we design the multi-objective optimisation reconstruction attacks based on the Nondominated Sorting Genetic Algorithm II (NSGA-II) \citep{deb2002fast}. NSGA-II is appraised for its elitism property - where the non-dominated individuals within a population remain unchanged in the next iteration. Consequently, this contributes to the speed-up of the evolutionary algorithm and the reduced computational complexity in finding the solutions to the multi-optimisation problem. Additionally, NSGA-II provides solutions that are spread across the objectives and are converged closer to the Pareto-optimal front. Arguably, while comparative studies tend to explore how state-of-the-art algorithms perform against NSGA-II \citep{zhou2011multiobjective, tian2021evolutionary}, we highlight that the efficiency of the NSGA-II algorithm might be appealing for an attacker that favours choosing a computationally inexpensive and a fairly accessible technique \citep{verma2021comprehensive}.

\subsubsection{Objectives}

The multi-objective optimisation problem is characterised by the formulation of objectives to be minimised in order to instigate effective reconstruction attacks. Based on the research questions in this work, we propose two objectives to minimise. 

The first objective $f_1$ is to minimise the distance of a selected synthetic sample to its nearest synthetic neighbours. As discussed earlier, samples that are in close vicinity to other datapoints could indicate that the generative model overfits on a given private datapoint. By optimising the distance objective, we aim to minimise the harmonic mean of distances to the neighbours in the population selection process. We can thereby represent the first objective for a given sample $d'_0$ as:

\begin{equation}
    f_1 = \bar{r}_{d'_0}
\end{equation}

The second objective $f_2$ is concerned with reducing the prediction error of the target variable of the query synthetic sample. A lower prediction error potentially corresponds to a higher certainty of the predictive features being closely aligned with the model training space. Meanwhile, a higher prediction error can generally indicate that a synthetic sample is not sufficiently realistic. The second objective can be denoted by:

\begin{equation}
    f_2 = \mathcal{L}\left(\hat{y'_0}, \mathcal{M}\left(x'_0\right)\right)
\end{equation}

where $\mathcal{L}$ is the cross-entropy loss for classification tasks or the root-mean-squared error for regression tasks. Therefore, in the dual-objective reconstruction attacks, the objective vector $F$ can be denoted as $F = \{f_1, f_2\}$.

While in principle it is possible to include constraints in the evolutionary multi-objective optimisation, we highlight that this is not strictly necessary for conducting reconstruction attacks for two reasons. First, we assume that an attacker possesses no domain knowledge of the constraints within the dataset and thus wants to launch attacks with minimum effort. Second, the use of GAN as the generative model allows the generation of synthetic samples without implicitly modelling the distributions or defining the correlation and constrained relationship features. Thus, since the GAN models already generate synthetic samples that do not necessarily meet the domain constraints, we refrain from introducing constraints in the evolutionary algorithm to avoid the significant alteration of the synthetic output.

\subsubsection{Generation Process}

The evolutionary algorithm is initialised by defining the number of generations $N_{gen}$ for evolutionary optimisation and producing a population $P$ consisting of $L$ individuals. While in principle there are several techniques for initialising the population $P$, we specify our initial population as $L$ replications of the query datapoint to be attacked. This is carried out because the query sample was already selected due to it is close vicinity to its neighbouring synthetic samples, thus, there is a greater confidence that the query datapoint is closer to the training space and therefore less perturbations are needed. 

%Second, the computational time to reach a valid final population is reduced since the initial sample is already fairly realistic and meets the categorical constraints, thus a Pareto-optimal front can be found with a lower $N_{gen}$.

In the survival phase, the NSGA-II algorithm evaluates the individuals of $P$ according to the objectives $F$. The non-dominance sorting ranks the individuals along multiple Pareto fronts. The individuals within the Pareto fronts are then further ranked according to their crowding distance, where individuals that are further away from the rest of the population in the objective space are ranked higher. Subsequently, the highest $L$ ranking individuals survive, while the other individuals are eliminated. 

In the binary tournament selection, samples of two individuals are drawn, the best individuals are selected as parents and are paired with parents selected from other tournament draws. The process ensures that the best parents are continuously selected for reproduction, while the remaining individuals are removed. The crossover phase generates new offspring from the features of the two parents. In the two-point crossover, the two crossover points between the parents are selected at random and the features are swapped to create the new features of the offspring.

Next, polynomial mutation is applied to randomly altering the features of the offspring from the crossover phase. All features have an equal likelihood of being altered. The mutation process contributes to the diversity of the produced population. The aforementioned steps are repeated until the predefined number of generations $N_{gen}$ is reached. The output of the previous steps is the final population from the final iteration.

\subsubsection{Multi-Criteria Decision Making (MCDM)}

The final population consists of $L$ individuals, all of which are ideally viable solutions to the multi-objective optimisation problem. To select a single solution as our reconstructed sample, we specify a weight vector $\Omega = \{ \omega_1, \omega_2 \}$ where $\omega_1, \omega_2$ are the weights corresponding to objectives $f_1, f_2$ respectively. We use the Achievement Scalarized Function (ASF) \citep{wierzbicki1980use} for applying the decomposition and finding the optimal sample from the Pareto-front solutions based on the specified weight vector $\Omega$. The evolutionary optimisation process is summarised in Algorithm \ref{alg:evomoe}.

%  While we defined two possible objectives for the optimisation problem, we place greater emphasis on $f_1$ for selecting a single solution out of all possible solutions in $P$ after the evolutionary optimisation is complete. To this end

\begin{algorithm}
\caption{Evolutionary multi-objective optimisation}
\label{alg:evomoe}
\begin{algorithmic}
\Require $d'_0 \in \mathcal{R} \,$ : query sample \newline 
    \hspace*{2em} $F=\{f_1, f_2\} \,$ : objectives \newline 
    %\hspace*{2em} constraints $G=\{g_1, \cdots, g_{m_{cat}} \}\,$,  \newline
    \hspace*{2em} $N_{gen} \,$: number of generations \newline
    \hspace*{2em} $L \,$: population size \newline
    \hspace*{2em} $\Omega = \{ \omega_1, \omega_2 \}$ : MCDM weight vector
\Ensure $\tilde{d}_0$ : reconstructed sample 
\State $P \gets init(d'_0, L) $
\For{$j=1$ to $j=N_{gen}$}
\State $P_{survive} \gets survive(P,F)$
\State $P_{parents} \gets binary\_tournament\_selection(P_{survive})$
\State $P_{offspring} \gets two\_point\_crossover(P_{parents})$
\State $P_{mutate} \gets mutate(P_{offspring})$
\State $P \gets P \cup P_{mutate}$
\EndFor
\State $\tilde{d}_0 \gets ASF(P, \Omega)$
\end{algorithmic}
\end{algorithm}

\section{Experimental Setup}

In this section, we present our framework's setup for conducting and analysing the privacy reconstruction attacks on synthetic datasets. 
%Our source code is publicly available \footnote{\url{https://github.com/abedshantti/}}.

\subsection{Datasets}
In this study, we consider the risk of privacy attacks on mixed-type tabular datasets. Tabular datasets are generally more susceptible to privacy attacks than high-resolution image datasets, as this is attributed to their lower dimensionality and the limited attributes of categorical features. Our focus in this work is mixed-type datasets that are typically used in the financial sector, however, the general notion can be applied to any tabular dataset in practice. All the datasets have a binary target feature, and thus are typically used for binary classification tasks. We preprocess the datasets to eliminate redundant features, samples with missing attributes and duplicate samples. The following datasets are used:

\begin{enumerate}
    \item \textbf{Adult} \citep{data_adult} is a popular tabular dataset that consists of a set of personal attributes, and the task is to predict whether each individual (sample) has an annual income greater than \$ 50,000.
    \item \textbf{Lending} \citep{data_lending} is a highly-dimensional dataset that encapsulates customers' financial history when applying for a loan, and the target feature is a binary indicator of whether the loan was repaid on time or not.
    \item \textbf{PAKDD} \citep{data_pakdd} is a customer risk scoring dataset that gives a binary credit risk rating to each anonymised customer of the Brazilian bank dataset based on the clients' personal and financial attributes.
    \item \textbf{Taiwan} \citep{data_taiwan} is a financial dataset where the aim is to predict whether customers will default on their loans based on their payment history and a few demographic attributes.
\end{enumerate}

A 50/50 split is carried out on the processed datasets to separate them into a training set for synthetic data generation and holdout split used only for evaluation. Table \ref{tb:datasets} presents a description of the datasets used for the synthetic data generation task.

\begin{table}[!htb]
\caption{Summary of the datasets used for privacy reconstruction attacks.}
\label{tb:datasets}
\centering
\small
%\resizebox{\columnwidth}{!}{
    \begin{tabular}{l l l l}
        \toprule
        Dataset &  Number of Samples & Continuous Features & Discrete Features  \\
        \midrule
        Adult & 24395 & 6 & 9  \\
        
        Lending  & 36954 & 53 & 8  \\
        
        PAKDD  & 22908 & 11 & 21   \\
        
        Taiwan  & 14982 & 20 & 4  \\
        
        \bottomrule
    \end{tabular}
%}
%}

\end{table}

\subsection{GAN Models}

We select three state-of-the-art tabular GAN models for generating synthetic data samples of the datasets. CTGAN \citep{xu2019modeling} is a Wasserstein-based gradient penalty loss GAN that utilises the selection of conditional vectors during training in order to ensure sufficient modelling of underrepresented discrete categories. Subsequently, CTGAN's synthetic output is highly realistic and the framework has been adapted for its applicability in other domains \citep{rosenblatt2020differentially,fang2022dp}. Meanwhile, CTAB-GAN \citep{zhao2021ctab} is a generative tabular GAN that improves the modelling of skewed numerical feature distributions and demonstrating a synthetic output that is well suitable for training machine learning models. CasTGAN \citep{alshantti2024castgan} adopts a cascaded architecture of generators, where each generator is assigned with the task of generating an individual feature. The model demonstrates significant improvement in modelling the correlations between data attributes and drastically reduces the number of invalid records given the constrained relationship between some data features. 

%While there exists some differentially private GANs, we purposefully choose to employ them in this work owing to two reasons. First, differentially private tabular generative model are not adapted the representation of discrete categories. The existing models use a combination of numeric and binary features, whilst discarding or transforming categorical columns which damages the semantic representation of the mixed-type tabular datasets. Second, it has been largely acknowledged that differentially private tabular GANs lead to unacceptable trade-offs to the quality of the synthetic output. It has been demonstrated that the quality of the private data is not suitable for training machine learning models or even representative of the real data, as differential privacy with large noise gradients destroys the statistical properties of the data, thus, deems it unusable.

\subsection{Evaluation Metrics}

The evaluation of privacy attacks on tabular data is a non-trivial task. Whereas in image reconstruction attacks it can be possible to evaluate the preciseness of privacy attacks by visually observing the output and verifying if resembles a real training sample, it is practically impossible to qualitatively assess the success rate of privacy attacks given the scale of dimensionality of the data. Instead, we use the following metrics to quantify the success rate of privacy attacks.

\textbf{Unique training samples}: As the reconstructions attacks are aimed at identifying 5\% of the training samples, it is undoubtedly desired by the attackers to target as many records as possible for maximising their benefit from the reconstruction attacks. A large diversity in the reconstructed samples entails that an attacker has high exposure to the different training samples and, thus, possess more knowledge for inflicting more harm.

\textbf{Hit rate}: The hit rate is a measure of how many training records were compromised by the privacy reconstruction attacks. Given that an attacker only targets the top 5\% records, we formulate the hit rate as the ratio of compromised samples divided by 5\% of the total records of the dataset. A compromised record is one where all the categorical features match those of the targeted training sample. Meanwhile, this is less straightforward for numerical features as they can take an infinite number of samples. Instead, we use the BIRCH clustering technique \citep{zhang1996birch} to divide a numeric feature into a number of categories determined by the clustering threshold which we set to $0.025$. Therefore, the definition of compromised numeric features applies if the reconstructed sample features belong to exactly the same clusters as the training record.

\textbf{Distance to closest record (DCR)}: The distance to closest record is a metric for quantifying how close a reconstructed sample is to the private training datapoint. For a given datapoint, a DCR=0 indicates that the reconstructed sample is an exact match of the training sample, and thus is considered a compromised datapoint. The data is both normalised and one-hot encoded for quantifying the DCR. From an attacker's point of view, the attacks should minimise the DCR in order to closely approach the training space.

Given that the attacks formulated in this study are black-box attacks, the attacker is in fact unable to quantify the success of their own attacks for devising new strategies that would improve the effectiveness of their attacks. Instead, the evaluation metrics are used by the defenders as a way for assessing the privacy robustness of their generative models. 

\section{Results}

\subsection{RQ1: The Impact of Access Levels}

We compare how different generative models access privileges contribute to the effectiveness of privacy attacks. Table \ref{tb:rq1_nunique} demonstrates the impact of the attack levels on the number of unique training samples targeted by those attacks. From the results, it is evident that Level I attacks, in which the attacker only has access to the synthetic data, evidently outperform Level II and Level III attacks in targeting a larger number of training samples. This is applicable for all the datasets and the three different GANs employed in the experimental analysis. Whereas, Level II and Level III attacks have a comparable performance in terms of attacking a less diverse population. The advantage for Level I attacks in targeting a more diverse subset of samples is attributed to the smaller size of the synthetic dataset in Level I attacks, in contrast to the large number of samples in Level II and Level III attacks stemming from the subsequent generation of synthetic samples. This utilisation of GANs in Level II and Level III attacks contributes to the generation of seemingly similar datapoints, causing the attacks to target a smaller subset of the feature space. 

\begin{table}[htb]
\small
\centering
\caption{Number of unique training samples attacked.}
\label{tb:rq1_nunique}
%\resizebox{\columnwidth}{!}{
\begin{tabular}{cc| ccc }
    \toprule
    & & \multicolumn{3}{c}{Unique Samples Attacked} \\
    & & CTGAN & CTAB-GAN & CasTGAN \\
    \midrule
    
    \multirow{3}{*}{Adult}
    & Level I & \textbf{932} & \textbf{851} & \textbf{1098} \\
    & Level II & 426 & 691 & 711  \\
    & Level III & 387 & 498 & 773 \\
    \hline
    
    \multirow{3}{*}{Lending} 
    & Level I & \textbf{1270} & \textbf{1426} & \textbf{1290}  \\
    & Level II & 544 & 1096 & 574  \\
    & Level III & 656 & 1121 & 877 \\
    \hline
    
    \multirow{3}{*}{PAKDD} 
    & Level I & \textbf{832} & \textbf{762} & \textbf{940}  \\
    & Level II & 357 & 503 & 594 \\
    & Level III & 521 & 413 & 680 \\
    \hline
    
    \multirow{3}{*}{Taiwan}
    & Level I & \textbf{601} & \textbf{573} & \textbf{483} \\
    & Level II & 275 & 367 & 327  \\
    & Level III & 305 & 249 & 333 \\
    \bottomrule
\end{tabular}
%}

\end{table}

Table \ref{tb:rq1_hit} outlines the hit rate of the attack levels on synthetic datasets generated by the different GAN models. It can be observed from Table \ref{tb:rq1_hit} that no attack strategy significantly dominated the other levels. Level III attacks achieved the highest hit rate on four synthetic datasets, while Level I and Level II attacks had the highest hit rate on two and three datasets respectively. Furthermore, it can be observed that no training records were compromised for the Lending dataset, due to its high dimensionality, demonstrating the challenge in  matching all the categories and the numerical attributes clusters for a large number of features. This also explains why a low hit rate was achieved on the PAKDD dataset for all the attack levels and the generative models.

\begin{table}[htb]
\small
\centering
\caption{Hit rate of reconstruction attacks types on GAN models. }
\label{tb:rq1_hit}
%\resizebox{\columnwidth}{!}{
\begin{tabular}{cc| ccc }
    \toprule
    & & \multicolumn{3}{c}{Hit Rate} \\
    & & CTGAN & CTAB-GAN & CasTGAN \\
    \midrule
    
    \multirow{3}{*}{Adult}
    & Level I & 0.336 & 0.525 & 0.415 \\
    & Level II & 0.537 & 0.514 & \textbf{0.680} \\
    & Level III & \textbf{0.645} & \textbf{0.715} & 0.614 \\
    \hline
    
    \multirow{3}{*}{Lending} 
    & Level I & 0.000 & 0.000 & 0.000 \\
    & Level II & 0.000 & 0.000 & 0.000 \\
    & Level III & 0.000 & 0.000 & 0.000 \\
    \hline
    
    \multirow{3}{*}{PAKDD} 
    & Level I & 0.047 & 0.026 & 0.063 \\
    & Level II & \textbf{0.102} & \textbf{0.035} & 0.094 \\
    & Level III & 0.091 & 0.024 & \textbf{0.108} \\
    \hline
    
    \multirow{3}{*}{Taiwan}
    & Level I & 0.757 & \textbf{0.797} & \textbf{0.754} \\
    & Level II & 0.630 & 0.661 & 0.589 \\
    & Level III & \textbf{0.761} & 0.730 & 0.578 \\
    \bottomrule
\end{tabular}
%}

\end{table}

Table \ref{tb:rq1_dcr} summarises the results of the average DCR of all the attack levels on the datasets. From the results, it can be observed that Level III attacks perform the best in terms of identifying the synthetic samples that are the closest to the training space. This is not surprising, as access to the trained generators for producing an indefinite amount of samples increases the probability of finding synthetic samples that are closer to the private training records. This gives rise to synthetic data records that in close vicinity of the synthetic data space, thus, higher certainty that the closely clustered samples correspond to an existing private datapoint. However, this also highlights the evident trade-off between the proximity to the training data space and the diversity of samples targeted, as it can be observed from Table \ref{tb:rq1_nunique} that the reconstructed samples correspond to a smaller number of training samples, hence several reconstructed synthetic datapoints attributed to the same training record. 

\begin{table}[htb]
\small
\centering
\caption{Distance to closest record (DCR) comparison between the different attack types on GAN models.}
\label{tb:rq1_dcr}
%\resizebox{\columnwidth}{!}{
\begin{tabular}{cc| ccc }
    \toprule
    & & \multicolumn{3}{c}{DCR} \\
    & & CTGAN & CTAB-GAN & CasTGAN \\
    \midrule
    
    \multirow{3}{*}{Adult}
    & Level I & 0.171 & 0.081 & 0.081 \\
    & Level II & 0.342 & 0.120 & 0.081 \\
    & Level III & \textbf{0.102} & \textbf{0.047} & \textbf{0.074} \\
    \hline
    
    \multirow{3}{*}{Lending} 
    & Level I & 0.597 & 0.559 & 0.674 \\
    & Level II & 0.845 & 0.462 & 0.848 \\
    & Level III & \textbf{0.471} & \textbf{0.450} & \textbf{0.645} \\
    \hline
    
    \multirow{3}{*}{PAKDD} 
    & Level I & 1.094 & 1.083 & 0.468 \\
    & Level II & 1.050 & \textbf{0.700} & \textbf{0.349} \\
    & Level III & \textbf{0.639} & 1.251 & 0.369 \\
    \hline
    
    \multirow{3}{*}{Taiwan}
    & Level I & 0.054 & 0.046 & \textbf{0.065} \\
    & Level II & \textbf{0.050} & 0.052 & 0.093 \\
    & Level III & 0.051 & \textbf{0.040} & 0.107 \\
    \bottomrule
\end{tabular}
%}

\end{table}

\subsection{RQ2: Selection Attacks Using Machine Learning Predictions}

We now investigate the impact of using the prediction error of the synthetic samples in addition to the harmonic mean for reconstructing training samples. The machine learning model is trained on the private data, and it is assumed that an attacker has black-box access to the API of the model; only able to query samples and obtain their prediction. For ranking the synthetic samples, a set of weight vectors is used in order to balance the ratios between the closeness to other synthetic samples, $\omega_1$, and the binary cross-entropy loss between the predictions and the target label of the synthetic data, $\omega_2$. Namely, for $\Omega = \{\omega_1, \omega_2\}$ we use $\Omega_1 = \{0.50, 0.50\}$, $\Omega_2 = \{0.75, 0.25\}$ and $\Omega_3 = \{1.00, 0.00\}$, where $\Omega_3$ is simply the reconstruction attacks relying only on the distance to the synthetic neighbours. 

In order to intuitively present the results of RQ2, we compare the performance of Level III attacks as they exhibited the smallest DCR, entailing a higher significance reconstruction threat to the original training space. In addition, we explore RQ2 using CTGAN since it is the most efficient GAN model to train, and also because the results in Tables \ref{tb:rq1_nunique}-\ref{tb:rq1_dcr} demonstrate that there is no major disparity between the GAN models, indicating that none of the employed GAN models are particularly more robust or susceptible to reconstruction attacks than the others. 

\begin{table}[htb]
\small
\centering
\caption{The effect of utilising machine learning predictions on the performance of reconstruction attacks.}
\label{tb:rq2}
%\resizebox{\columnwidth}{!}{
\begin{tabular}{cc| ccc }
    \toprule
    & & Unique Samples & Hit Rate &DCR \\
    \midrule
    
    \multirow{3}{*}{Adult}
    & $\Omega_1$ & 287 & \textbf{0.693} & \textbf{0.087} \\
    & $\Omega_2$ & 297 & 0.679 & 0.088 \\
    & $\Omega_3$ & \textbf{387} & 0.645 & 0.102 \\
    \hline
    
    \multirow{3}{*}{Lending} 
    & $\Omega_1$ & 596 & 0.000 & 0.528 \\
    & $\Omega_2$ & 594 & 0.000 & 0.483 \\
    & $\Omega_3$ & \textbf{656} & 0.000 & \textbf{0.471} \\
    \hline
    
    \multirow{3}{*}{PAKDD} 
    & $\Omega_1$ & 484 & \textbf{0.109} & 0.639 \\
    & $\Omega_2$ & 499 & 0.099 & \textbf{0.625} \\
    & $\Omega_3$ & \textbf{521} & 0.091 & 0.639 \\
    \hline
    
    \multirow{3}{*}{Taiwan}
    & $\Omega_1$ & \textbf{550} & 0.689 & 0.057 \\
    & $\Omega_2$ & 393 & \textbf{0.801} & \textbf{0.042} \\
    & $\Omega_3$ & 305 & 0.761 & 0.051 \\
    \bottomrule
\end{tabular}
%}

\end{table}

The impact of utilising the prediction loss on the reconstruction attacks can be observed in Table \ref{tb:rq2}. It can be noticed that in general that the most diverse targeting of training samples is attained by $\Omega_3$. Meanwhile, it appears that relying on the BCE loss for attacking the synthetic samples leads to an increased hit rate. On the other hand, we can observe that DCR is highly dependent on the size of the datasets, where assigning equal weights for the proximity and the prediction losses appears to strongly impact lower dimensional datasets. As the difference in the Table \ref{tb:rq2} results is marginal, it can be deduced that there is little evidence to support that the utilisation of a discriminative model contributes to the the increased effectiveness of reconstruction attacks. 

\subsection{RQ3: The Role of Evolutionary Multi-Objective Optimisation}

We now study the impact of multi-objective optimisation on the performance of reconstruction attacks. As opposed to the experiments for answering RQ2, we assume that in this scenario the attacker has no access to the machine learning model learnt from the training set, but rather trains a new prediction model on the synthetic dataset. In contrast to the attacking scenarios in RQ1 and RQ2, reconstruction attacks using multi-objective optimisation actively perturb the data attributes during the evolutionary process, and thus, the synthetic datapoints have the potential to move closer to or further away from their corresponding private training datapoints. Since there was no evident advantage for weight vectors in RQ2, we further devise reconstruction attacks using the same three weight vectors in the multi criteria decision making stage after the final population has been generated in order to outline the difference in the precision of the attacks. Table \ref{tb:rq3} demonstrates these results. 

\begin{table}[htb]
\small
\centering
\caption{Comparison of reconstruction attacks performance of evolutionary multi-objective optimisation and non-evolutionary $\Omega_3$ attacks in RQ2 (w/o MoE).}
\label{tb:rq3}
%\resizebox{\columnwidth}{!}{
\begin{tabular}{cc| ccc }
    \toprule
    & & Unique Samples & Hit Rate &DCR \\
    \midrule
    
    \multirow{4}{*}{Adult}
    & w/o MoE & \textbf{387} & 0.645 & 0.102 \\
    & $\Omega_1$ & 290 & 0.233 & 0.281 \\
    & $\Omega_2$ & 277 & 0.267 & 0.194 \\
    & $\Omega_3$ & 383 & \textbf{0.679} & \textbf{0.101} \\
    \hline
    
    \multirow{4}{*}{Lending} 
    & w/o MoE & \textbf{656} & 0.000 & 0.471 \\
    & $\Omega_1$ & 496 & 0.000 & 0.481 \\
    & $\Omega_2$ & 521 & 0.000 & 0.453 \\
    & $\Omega_3$ & 559 & 0.000 & \textbf{0.446} \\
    \hline
    
    \multirow{4}{*}{PAKDD} 
    & w/o MoE & \textbf{521} & 0.091 & 0.639 \\
    & $\Omega_1$ & 462 & 0.098 & 0.908 \\
    & $\Omega_2$ & 425 & \textbf{0.121} & 0.652 \\
    & $\Omega_3$ & 508 & 0.095 & \textbf{0.638} \\
    \hline
    
    \multirow{4}{*}{Taiwan}
    & w/o MoE & \textbf{305} & 0.761 & 0.051 \\
    & $\Omega_1$ & 254 & 0.583 & 0.062 \\
    & $\Omega_2$ & 275 & 0.698 & 0.058 \\
    & $\Omega_3$ & 279 & \textbf{0.769} & \textbf{0.050} \\
    \bottomrule
\end{tabular}
%}

\end{table}

There are multiple observations that can be made from Table \ref{tb:rq3}. First, it can be noticed that allocating a higher weight for the BCE loss in order to select samples from the Pareto solutions that optimise both objectives lead to a drop in the success rate of the attacks. It is clear that in most cases $\Omega_1$ and $\Omega_2$ perform worse in targeting diverse samples and perturbing the synthetic samples closer to the training space. Meanwhile, there are noticeable trade-offs between not using evolutionary algorithms and using NSGA-II for perturbations and nevertheless selecting the optimal solutions that only minimises the distance to the synthetic neighbours. It is evident that the use of evolutionary algorithms to perturb synthetic samples reduces the diversity of the targeted training samples, even when using $\Omega_3$. On the other hand, we notice that evolutionary algorithms and MCDM using $\Omega_3$ moderately increases the precision of the reconstruction attacks as observed from the hit rate and the DCR.

\subsection{Against Benchmark Attacks}

We further compare the performance of our attacks against the attacks devised by \cite{hayes2019logan} and \cite{chen2020gan}. While both works formulated the attacks as membership inference attacks, we re-adapt the attacks for the re-identification scenario without a holdout set. For emulating the full black-box LOGAN attack in \citep{hayes2019logan}, we first train a surrogate CTAB-GAN model on the available synthetic data to generate a number of synthetic samples. We then train another CTAB-GAN model on the newly generated synthetic samples. We then query the synthetic data samples from the CTGAN model on the CTAB-GAN model and rank the samples according to the output of the discriminator. Subsequently, the top 5\% samples are selected as the re-identified samples. The full implementation of the full black-box LOGAN attacks can be found in the original paper \citep{hayes2019logan}. 

In addition, we implement two variants of the GAN-Leaks attacks proposed by \cite{chen2020gan}. We first consider the base full black-box attacks in which $1$-nearest neighbour is used to rank the synthetic samples according to the distance to their neighbours. The top 5\% samples closest to their neighbours are selected as the candidate samples. Meanwhile, we also consider the calibrated GAN-Leaks attacks which consider the training of a second surrogate model and computing the membership calibrated error between the surrogate and the original GAN model. We refer the reader to the original paper \citep{chen2020gan} for the full implementation details. We compare our reconstruction attacks against the baseline attacks and demonstrate the results in Table \ref{tb:baseline_atts}.

\begin{table}[htb]
\small
\centering
\caption{Comparing the number of unique samples, hit rate and DCR of our multi-objective optimised $\Omega_3$ attacks against baseline GAN attacks.}
\label{tb:baseline_atts}
%\resizebox{\columnwidth}{!}{
\begin{tabular}{cc| ccc }
    \toprule
    & & Unique Samples & Hit Rate & DCR \\
    \midrule
    
    \multirow{4}{*}{Adult}
    & LOGAN & \textbf{827} & 0.019 & 1.429 \\
    & GAN-Leaks & 804 & 0.489 & 0.128 \\
    & GAN-Leaks Calibrated & 766 & 0.104 & 0.570 \\
    & Ours & 383 & \textbf{0.679} & \textbf{0.101} \\
    \hline
    
    \multirow{4}{*}{Lending} 
    & LOGAN & \textbf{1701} & 0.000 & 1.073 \\
    & GAN-Leaks & 1224 & 0.000 & 0.543 \\
    & GAN-Leaks Calibrated & 1362 & 0.000 & 0.996 \\
    & Ours & 559 & 0.000 & \textbf{0.446} \\
    \hline
    
    \multirow{4}{*}{PAKDD} 
    & LOGAN & 757 & 0.023 & 1.766 \\
    & GAN-Leaks & 803 & 0.052 & 0.693 \\
    & GAN-Leaks Calibrated & \textbf{835} & 0.034 & 1.205 \\
    & Ours & 508 & \textbf{0.095} & \textbf{0.638} \\
    \hline
    
    \multirow{4}{*}{Taiwan}
    & LOGAN & \textbf{657} & 0.594 & 0.139 \\
    & GAN-Leaks & 493 & \textbf{0.844} & \textbf{0.042} \\
    & GAN-Leaks Calibrated & 558 & 0.393 & 0.203 \\
    & Ours & 279 & 0.769 & 0.050 \\
    \bottomrule
\end{tabular}
%}

\end{table}

Inspecting Table \ref{tb:baseline_atts} demonstrates that the LOGAN attacks achieved the highest diversity in the number of training samples targeted. Querying on the discriminator is minimally deterministic as the output of the discriminator depends on various factors during the training process. As such, we notice that contrastingly the LOGAN achieves the worst hit rate and DCR scores amongst the other approaches. Meanwhile, we observe that the calibrated GAN-leaks attack achieve a lower success rate than the base GAN-Leaks attacks. This can be attributed to the fact that the calibrated attacks are well designed for the membership inference attacks as they aim to estimate the membership probability, thus, they are significantly less effective in our case, where no holdout set is available for querying. Finally, we notice that the base GAN-Leaks attacks achieve the closest performance to our reconstruction attacks. This indeed validates that the use of proximity measures for selecting the samples according to their vicinity to their neighbours can be a good indicator of overfitting or potential data leakage. Nevertheless, we observe that the reconstruction attacks we proposed are the most successful attacks as they achieve the best hit rate and DCR scores on most of the datasets. However, we highlight that this comes at the expense of the reduced number of targeted training samples. 

\subsection{Defence Mechanisms}

While it is common knowledge that differential privacy yields unacceptable trade-offs between the privacy and utility of the synthetic data \citep{shokri2015privacy}, it is nevertheless interesting to analyse the robustness of differential privacy against our evolutionary multi-objective attacks. For conducting this analysis, we employ DP-auto-GAN \citep{tantipongpipat2021differentially} for generating differentially private synthetic samples. While there is no scarcity of differentially private tabular GAN models in the literature, we employ DP-auto-GAN as it achieved better performance than some of the existing benchmarks, and for it is ability to handled mixed-type datasets. A more thorough coverage of differentially private GANs can be found in Section 6. 

For generating differentially private synthetic samples, we use the same parameters for DP-auto-GAN as in the original paper in \citep{tantipongpipat2021differentially}, where a lower privacy budget parameter $\epsilon$ increases the privacy of the synthetic data, while a higher $\epsilon$ produces less rigid privacy guarantees. The results in Table \ref{tb:defence_privacy} demonstrate this behaviour on the Adult dataset. We notice that all differentially private models reduced the number of training samples targeted by reconstruction attacks, which reduces the risk of exposure of  training datapoints. Moreover, the hit rate has been significantly reduced against our reconstruction attacks, hence, a fewer number of records suffering from leakage. It can also be observed how the DCR grows considerably with stricter privacy guarantees, thus demonstrating the disparity between the training set and the differentially private synthetic datasets.

\begin{table}[htb]
\small
\centering
\caption{Comparison of differential privacy against our model - attack success}
\label{tb:defence_privacy}
%\resizebox{\columnwidth}{!}{
\begin{tabular}{c| ccc }
    \toprule
    &  Unique Samples & Hit Rate & DCR \\
    \midrule
    
    $\epsilon = 0.36$ &  96 & 0.049 & 0.638 \\
    $\epsilon = 0.51$ &  78 & 0.024 & 0.381 \\
    $\epsilon = 1.01$ & 101 & 0.071 & 0.103 \\
    CTGAN             & 383 & 0.679 & 0.101 \\
    
    \bottomrule
\end{tabular}
%}

\end{table}

For evaluating the utility of the synthetic datasets, we consider some of the metrics presented in \citep{alshantti2024castgan}. We measure the F1-score of the predictions on the test data using models trained on the synthetic datasets. In addition, the Kolmogrov-Smirnov two-sample test score (KS-statistic) is used to quantify the univariate distribution statistical errors of the features of the synthetic output in comparison to the private training set. Meanwhile, the correlation Root Mean Squared Error (correlation RMSE) measures the error of the attribute correlations between the real data and the synthetic data. The results are reported in Table \ref{tb:defence_utility}. From the results, we can observe how the differential privacy significantly diminishes the quality of the data. We can further notice from the F1-score in Table \ref{tb:defence_utility} that the synthetic datasets produced by DP-auto-GAN are significantly less suitable for training predictive errors due to the large error in predicting the target feature. Moreover, the structure of the data is significantly distorted, which is reflected by the high KS-statistic and correlation RMSE measures. Therefore, it is obvious that using differential privacy against our reconstruction attacks comes at the expense of generating synthetic output that represents the properties of the original data.

\begin{table}[htb]
\small
\centering
\caption{Comparison of differential privacy against our model - data utility}
\label{tb:defence_utility}
%\resizebox{\columnwidth}{!}{
\begin{tabular}{c| ccc }
    \toprule
    &  F1-score & KS-statistic & Correlation RMSE \\
    \midrule
    
    $\epsilon = 0.36$ &  0.177 & 0.220 & 0.288 \\
    $\epsilon = 0.51$ &  0.331 & 0.191 & 0.230 \\
    $\epsilon = 1.01$ &  0.263 & 0.199 & 0.234 \\
    CTGAN             &  0.624 & 0.150 & 0.057 \\
    
    \bottomrule
\end{tabular}
%}

\end{table}

\section{Related Works}

% Maybe merge this section with sec2

% Comment about attacks

% Comment about defences

% Some GAN models -- talk about limitations of some privacy models

% Evaluation is extremely difficult, expert input might be needed to judge on what constitutes a breach

% Attack models .. not much done, very strong assumptions hold

\subsection{Privacy Attacks}

While privacy attacks are adequately studied for discriminative models, they have been explored to a much lesser extent in the generative model literature. \cite{hayes2019logan} proposed membership attacks on GANs in the white-box and black-box settings. The attacks work by querying samples on the discriminator and attributing the samples with higher discriminator scores as training samples. Unsurprisingly, it was demonstrated that white-box attacks that query samples on the discriminator are more successful than attacks that were evaluated on GANs trained on a surrogate dataset. \cite{hilprecht2019monte} proposed the two types of membership inference attacks: Monte Carlo attacks on generative models and Reconstruction attacks on VAEs. The experimental results demonstrated that the proposed Monte Carlo attacks outperformed existing black-box membership inference attacks on generative models, while showing that reconstruction attacks on the VAEs were extremely effective, which indicates that variational autoencoders are considerably susceptible to overfitting. Moreover, \cite{chen2020gan} also devised membership inference attacks against GANs in both knowledgeable white-box settings and less knowledgeable black-box settings and produced findings consistent to previous studies while outperforming existing membership inference attacks. Similarly, \cite{zhang2022membership} comprehensively studied the impact of membership inference attacks against longitudinal health data with no assumptions about the structure of the generative model used, and found that partially synthetic data is significantly more prone to inference attacks than fully synthetic data. Meanwhile, \cite{hu2021tablegan} proposed membership collision attacks on generative models by training multiple shadow generative models and attributing the overlapping samples from the shadow models as training samples. However, it is observed that this type of attack is unsuccessful against mixed-type tabular datasets, where the data is discretised to ensure that the overlap of samples takes place. It is worth noting that in the aforementioned works, the success rate of membership inference attacks is reported to significantly drop as the size of the training set increases. Therefore, the attacks are only slightly better than random guessing the query samples when the large datasets were used for training. Additionally, the majority of the literature on generative adversarial attacks is dedicated to computer vision application and image data, while insufficiently addressing the tabular data domain.

\subsection{Privacy Defences}

On the other hand, mitigation mechanisms for generative models have been adequately explored in the literature. \cite{zhang2017privbayes} presented PrivBayes as a Bayesian network-based generative model to create synthetic data while conforming to differential privacy guarantees. However, it has been demonstrated by \cite{xu2019modeling} that the data generated by PrivBayes exhibits much lower data utility for machine learning tasks than the GAN counterparts. \cite{xie2018differentially} proposed a differentially private GAN that provides privacy guarantees with respect to the training data by clipping the weights and adding noise to the gradients. \cite{jordon2018pate} devised PATE-GAN, which is a generative framework adopting a teacher-student ensemble method to ensure strict privacy guarantees and improved quality of the synthetic output. \cite{chen2020gs} proposed a differentially private GAN model that provides rigid privacy guarantees by sanitising the generator's output, while maintaining optimal training of the discriminator. \cite{lee2021invertible} designed a generative model consisting of a GAN and a VAE combined with the negative log density regularisation for adjusting the trade-off between the privacy and utility of synthetic tabular data. The combination of VAE and GAN was also implemented by \cite{torfi2022differentially} where an auto-encoder was combined with a convolutional GAN for achieving differential privacy for tabular data synthesis. PAR-GAN is a framework that composes of a single generator and multiple discriminator for training the GAN on multiple disjoint partitions of the training data in order to generalise on the data and avoid the memorisation of individual samples \citep{chen2021pargan}. Meanwhile, Fed-Avg GAN was proposed by \cite{mcmahan2017learning} for synthesis of decentralised data in federated learning, while providing different privacy guarantees at a user-level. In the aforementioned works, it is evident that the quality of the synthetic data is considerably sacrificed in order to ensure the privacy of the training data. \cite{ganev2022robin} conducted a study for investigating the distortion of class distributions in the synthetic datasets induced by several differential privacy synthesis techniques. Notwithstanding, \cite{lu2019empirical} argue that the strict differential privacy guarantee is a concept that has been thoroughly explored in the research community, whereas, it might be sufficient for industries in practice to rely on the GAN-based synthetic output for legal purposes and improved data quality. 

%Moreover, \cite{dankar2012application} assert that differential privacy as a theoretical framework is optimally crafted for functions rather than data.

\subsection{Privacy Risk Evaluation}

In principle, quantifying the threat on training data by privacy attacks is an intrinsically challenging task. \cite{song2021systematic} proposed a privacy risk score based on the Bayes' theorem to quantify the probability of a query sample being a member of the training for evaluating the success membership inference attacks. Whereas, \cite{park2014pegs} formulated the disclosure risk score as a measure to estimate the likelihood of recovering a feature of a record given the attacker has knowledge of the other features for the record. Meanwhile, \cite{chen2021pargan} used the generalisation gap between the discriminator scores of the training set and holdout set to quantify how well does differential privacy succeeds in reducing the gap. Qualitative analysis of synthetic data is a widely adopted approach in the image synthesis where the human eye can be used for verifying whether the identity of real individuals or target classes can be determined from the synthetically generated output \citep{,hitaj2017deep, tseng2020compressive, wang2021variational}. Meanwhile, \cite{aivodji2019gamin} relied on human interpretation of the synthetic output by conducting a survey and asking the participants to guess the labels of the synthetic images that were presented to them. The privacy risk on tabular data remains considerably more challenging to quantify than for visual applications.

% Attack Evals: qualitative analysis not possible, can be subjective, sometimes experts were used

\section{Conclusion}

In this work, we designed and implemented reconstruction attacks on tabular synthetic data. Traditionally, the vulnerability of synthetic data is predominantly studied in the context of membership inference attacks; where an attacker with a query set consisting of private training records and synthetic records tries to infer which records were used for training a generative model. We note that while the study of membership inference attacks emphasise the need for better defending mechanisms for generative models, membership inference attacks are built on the strong and potentially unreasonable assumption that an attacker already has a query set that comprises of private samples. Instead, we shed light on the case where an attacker tries to recover training data records via the synthetic data and possibly black-box access to the generative and discriminative models trained on the original data.

Our experimental results demonstrate that reconstruction attacks pose a major threat in recovering sensitive information in the training set. In accordance with intuition, our results demonstrated that access to the generator by an attacker for producing a large number of synthetic samples yields more precise reconstructions of the private datapoints. In addition, the use of multi-objective optimisation and evolutionary algorithms enable the perturbation of synthetic samples for reconstructing training samples more effectively. Nevertheless, we highlight that the evaluation of privacy attacks on mixed-type tabular data is a non-trivial task. Furthermore, the current mitigation techniques for hindering privacy attacks are significantly detrimental to the quality and the utility of the synthetic data. We therefore aim that the introduction of reconstruction attacks in the synthetic tabular domain motivates for more robust defence mechanisms that guarantee privacy whilst not jeopardising the data's usability.

\section*{Acknowledgments}
This work was supported by DNB ASA through the funding of this research project. The authors would like to thank Damiano Varagnolo for his insights and valuable contributions.

%Bibliography
\bibliographystyle{apalike}  
\bibliography{ref}  

\begin{thebibliography}{}

\bibitem[Abadi et~al., 2016]{abadi2016deep}
Abadi, M., Chu, A., Goodfellow, I., McMahan, H.~B., Mironov, I., Talwar, K., and Zhang, L. (2016).
\newblock Deep learning with differential privacy.
\newblock In {\em Proceedings of the 2016 ACM SIGSAC conference on computer and communications security}, pages 308--318.

\bibitem[Agrawal and Srikant, 2000]{agrawal2000privacy}
Agrawal, R. and Srikant, R. (2000).
\newblock Privacy-preserving data mining.
\newblock In {\em Proceedings of the 2000 ACM SIGMOD international conference on Management of data}, pages 439--450.

\bibitem[A{\"\i}vodji et~al., 2019]{aivodji2019gamin}
A{\"\i}vodji, U., Gambs, S., and Ther, T. (2019).
\newblock Gamin: An adversarial approach to black-box model inversion.
\newblock {\em arXiv preprint arXiv:1909.11835}.

\bibitem[Alshantti et~al., 2024]{alshantti2024castgan}
Alshantti, A., Varagnolo, D., Rasheed, A., Rahmati, A., and Westad, F. (2024).
\newblock Castgan: Cascaded generative adversarial network for realistic tabular data synthesis.
\newblock {\em IEEE Access}.

\bibitem[Cai et~al., 2021]{cai2021generative}
Cai, Z., Xiong, Z., Xu, H., Wang, P., Li, W., and Pan, Y. (2021).
\newblock Generative adversarial networks: A survey toward private and secure applications.
\newblock {\em ACM Computing Surveys (CSUR)}, 54(6):1--38.

\bibitem[Chen et~al., 2020a]{chen2020gs}
Chen, D., Orekondy, T., and Fritz, M. (2020a).
\newblock Gs-wgan: A gradient-sanitized approach for learning differentially private generators.
\newblock {\em Advances in Neural Information Processing Systems}, 33:12673--12684.

\bibitem[Chen et~al., 2020b]{chen2020gan}
Chen, D., Yu, N., Zhang, Y., and Fritz, M. (2020b).
\newblock Gan-leaks: A taxonomy of membership inference attacks against generative models.
\newblock In {\em Proceedings of the 2020 ACM SIGSAC conference on computer and communications security}, pages 343--362.

\bibitem[Chen et~al., 2021]{chen2021pargan}
Chen, J., Wang, W.~H., Gao, H., and Shi, X. (2021).
\newblock Par-gan: Improving the generalization of generative adversarial networks against membership inference attacks.
\newblock In {\em Proceedings of the 27th ACM SIGKDD Conference on Knowledge Discovery \& Data Mining}, pages 127--137.

\bibitem[Choi et~al., 2017]{choi2017generating}
Choi, E., Biswal, S., Malin, B., Duke, J., Stewart, W.~F., and Sun, J. (2017).
\newblock Generating multi-label discrete patient records using generative adversarial networks.
\newblock In {\em Machine learning for healthcare conference}, pages 286--305. PMLR.

\bibitem[Deb et~al., 2002]{deb2002fast}
Deb, K., Pratap, A., Agarwal, S., and Meyarivan, T. (2002).
\newblock A fast and elitist multiobjective genetic algorithm: Nsga-ii.
\newblock {\em IEEE transactions on evolutionary computation}, 6(2):182--197.

\bibitem[Dwork, 2008]{dwork2008differential}
Dwork, C. (2008).
\newblock Differential privacy: A survey of results.
\newblock In {\em International conference on theory and applications of models of computation}, pages 1--19. Springer.

\bibitem[Engelmann and Lessmann, 2021]{engelmann2021conditional}
Engelmann, J. and Lessmann, S. (2021).
\newblock Conditional wasserstein gan-based oversampling of tabular data for imbalanced learning.
\newblock {\em Expert Systems with Applications}, 174:114582.

\bibitem[Esteban et~al., 2017]{esteban2017real}
Esteban, C., Hyland, S.~L., and R{\"a}tsch, G. (2017).
\newblock Real-valued (medical) time series generation with recurrent conditional gans.
\newblock {\em arXiv preprint arXiv:1706.02633}.

\bibitem[Fang et~al., 2022]{fang2022dp}
Fang, M.~L., Dhami, D.~S., and Kersting, K. (2022).
\newblock Dp-ctgan: Differentially private medical data generation using ctgans.
\newblock In {\em International Conference on Artificial Intelligence in Medicine}, pages 178--188. Springer.

\bibitem[Fredrikson et~al., 2014]{fredrikson2014privacy}
Fredrikson, M., Lantz, E., Jha, S., Lin, S., Page, D., and Ristenpart, T. (2014).
\newblock Privacy in pharmacogenetics: An $\{$End-to-End$\}$ case study of personalized warfarin dosing.
\newblock In {\em 23rd USENIX Security Symposium (USENIX Security 14)}, pages 17--32.

\bibitem[Ganev et~al., 2022]{ganev2022robin}
Ganev, G., Oprisanu, B., and De~Cristofaro, E. (2022).
\newblock Robin hood and matthew effects: Differential privacy has disparate impact on synthetic data.
\newblock In {\em International Conference on Machine Learning}, pages 6944--6959. PMLR.

\bibitem[Gondara and Wang, 2018]{gondara2018mida}
Gondara, L. and Wang, K. (2018).
\newblock Mida: Multiple imputation using denoising autoencoders.
\newblock In {\em Pacific-Asia conference on knowledge discovery and data mining}, pages 260--272. Springer.

\bibitem[Goodfellow et~al., 2014]{goodfellow2014generative}
Goodfellow, I., Pouget-Abadie, J., Mirza, M., Xu, B., Warde-Farley, D., Ozair, S., Courville, A., and Bengio, Y. (2014).
\newblock Generative adversarial nets.
\newblock {\em Advances in neural information processing systems}, 27.

\bibitem[Hayes et~al., 2019]{hayes2019logan}
Hayes, J., Melis, L., Danezis, G., and De~Cristofaro, E. (2019).
\newblock Logan: Membership inference attacks against generative models.
\newblock In {\em Proceedings on Privacy Enhancing Technologies (PoPETs)}, volume 2019, pages 133--152. De Gruyter.

\bibitem[Hidano et~al., 2017]{hidano2017model}
Hidano, S., Murakami, T., Katsumata, S., Kiyomoto, S., and Hanaoka, G. (2017).
\newblock Model inversion attacks for prediction systems: Without knowledge of non-sensitive attributes.
\newblock In {\em 2017 15th Annual Conference on Privacy, Security and Trust (PST)}, pages 115--11509. IEEE.

\bibitem[Hilprecht et~al., 2019]{hilprecht2019monte}
Hilprecht, B., H{\"a}rterich, M., and Bernau, D. (2019).
\newblock Monte carlo and reconstruction membership inference attacks against generative models.
\newblock {\em Proceedings on Privacy Enhancing Technologies}, 2019(4):232--249.

\bibitem[Hitaj et~al., 2017]{hitaj2017deep}
Hitaj, B., Ateniese, G., and Perez-Cruz, F. (2017).
\newblock Deep models under the gan: information leakage from collaborative deep learning.
\newblock In {\em Proceedings of the 2017 ACM SIGSAC conference on computer and communications security}, pages 603--618.

\bibitem[Hu et~al., 2021]{hu2021tablegan}
Hu, A., Xie, R., Lu, Z., Hu, A., and Xue, M. (2021).
\newblock Tablegan-mca: Evaluating membership collisions of gan-synthesized tabular data releasing.
\newblock In {\em Proceedings of the 2021 ACM SIGSAC Conference on Computer and Communications Security}, pages 2096--2112.

\bibitem[Ishibuchi et~al., 2016]{ishibuchi2016performance}
Ishibuchi, H., Imada, R., Setoguchi, Y., and Nojima, Y. (2016).
\newblock Performance comparison of nsga-ii and nsga-iii on various many-objective test problems.
\newblock In {\em 2016 IEEE Congress on Evolutionary Computation (CEC)}, pages 3045--3052. IEEE.

\bibitem[Jayaraman and Evans, 2019]{jayaraman2019evaluating}
Jayaraman, B. and Evans, D. (2019).
\newblock Evaluating differentially private machine learning in practice.
\newblock In {\em 28th USENIX Security Symposium (USENIX Security 19)}, pages 1895--1912.

\bibitem[Jia et~al., 2019]{jia2019memguard}
Jia, J., Salem, A., Backes, M., Zhang, Y., and Gong, N.~Z. (2019).
\newblock Memguard: Defending against black-box membership inference attacks via adversarial examples.
\newblock In {\em Proceedings of the 2019 ACM SIGSAC conference on computer and communications security}, pages 259--274.

\bibitem[Jordon et~al., 2018]{jordon2018pate}
Jordon, J., Yoon, J., and Van Der~Schaar, M. (2018).
\newblock Pate-gan: Generating synthetic data with differential privacy guarantees.
\newblock In {\em International conference on learning representations}.

\bibitem[Karras et~al., 2019]{karras2019style}
Karras, T., Laine, S., and Aila, T. (2019).
\newblock A style-based generator architecture for generative adversarial networks.
\newblock In {\em Proceedings of the IEEE/CVF conference on computer vision and pattern recognition}, pages 4401--4410.

\bibitem[Kingma et~al., 2014]{kingma2014semi}
Kingma, D.~P., Mohamed, S., Jimenez~Rezende, D., and Welling, M. (2014).
\newblock Semi-supervised learning with deep generative models.
\newblock {\em Advances in neural information processing systems}, 27.

\bibitem[Kohavi and Becker, 1996]{data_adult}
Kohavi, R. and Becker, B. (1996).
\newblock Adult data set.
\newblock \url{https://archive.ics.uci.edu/ml/datasets/adult}.

\bibitem[Koller and Friedman, 2009]{koller2009probabilistic}
Koller, D. and Friedman, N. (2009).
\newblock {\em Probabilistic graphical models: principles and techniques}.
\newblock MIT press.

\bibitem[Lee et~al., 2021]{lee2021invertible}
Lee, J., Hyeong, J., Jeon, J., Park, N., and Cho, J. (2021).
\newblock Invertible tabular gans: Killing two birds with one stone for tabular data synthesis.
\newblock {\em Advances in Neural Information Processing Systems}, 34:4263--4273.

\bibitem[{Lending Club}, 2018]{data_lending}
{Lending Club} (2018).
\newblock Loan default dataset.
\newblock \url{https://www.lendingclub.com/}.

\bibitem[Li et~al., 2006]{li2006t}
Li, N., Li, T., and Venkatasubramanian, S. (2006).
\newblock t-closeness: Privacy beyond k-anonymity and l-diversity.
\newblock In {\em 2007 IEEE 23rd international conference on data engineering}, pages 106--115. IEEE.

\bibitem[Liu et~al., 2019]{liu2019performing}
Liu, K.~S., Xiao, C., Li, B., and Gao, J. (2019).
\newblock Performing co-membership attacks against deep generative models.
\newblock In {\em 2019 IEEE International Conference on Data Mining (ICDM)}, pages 459--467. IEEE.

\bibitem[Long et~al., 2018]{long2018understanding}
Long, Y., Bindschaedler, V., Wang, L., Bu, D., Wang, X., Tang, H., Gunter, C.~A., and Chen, K. (2018).
\newblock Understanding membership inferences on well-generalized learning models.
\newblock {\em arXiv preprint arXiv:1802.04889}.

\bibitem[Lu et~al., 2019]{lu2019empirical}
Lu, P.-H., Wang, P.-C., and Yu, C.-M. (2019).
\newblock Empirical evaluation on synthetic data generation with generative adversarial network.
\newblock In {\em Proceedings of the 9th International Conference on Web Intelligence, Mining and Semantics}, pages 1--6.

\bibitem[McMahan et~al., 2017]{mcmahan2017learning}
McMahan, H.~B., Ramage, D., Talwar, K., and Zhang, L. (2017).
\newblock Learning differentially private recurrent language models.
\newblock {\em arXiv preprint arXiv:1710.06963}.

\bibitem[Melis et~al., 2019]{melis2019exploiting}
Melis, L., Song, C., De~Cristofaro, E., and Shmatikov, V. (2019).
\newblock Exploiting unintended feature leakage in collaborative learning.
\newblock In {\em 2019 IEEE Symposium on Security and Privacy (SP)}, pages 691--706. IEEE.

\bibitem[Narayanan and Shmatikov, 2008]{narayanan2008robust}
Narayanan, A. and Shmatikov, V. (2008).
\newblock Robust de-anonymization of large sparse datasets.
\newblock In {\em 2008 IEEE Symposium on Security and Privacy (sp 2008)}, pages 111--125. IEEE.

\bibitem[Nasr et~al., 2018]{nasr2018machine}
Nasr, M., Shokri, R., and Houmansadr, A. (2018).
\newblock Machine learning with membership privacy using adversarial regularization.
\newblock In {\em Proceedings of the 2018 ACM SIGSAC conference on computer and communications security}, pages 634--646.

\bibitem[Nasr et~al., 2019]{nasr2019comprehensive}
Nasr, M., Shokri, R., and Houmansadr, A. (2019).
\newblock Comprehensive privacy analysis of deep learning: Passive and active white-box inference attacks against centralized and federated learning.
\newblock In {\em 2019 IEEE symposium on security and privacy (SP)}, pages 739--753. IEEE.

\bibitem[{PAKDD}, 2009]{data_pakdd}
{PAKDD} (2009).
\newblock Pakdd 2009 data mining competition.
\newblock \url{https://pakdd.org/archive/pakdd2009/front/show/competition.html}.

\bibitem[Park et~al., 2018]{park2018data}
Park, N., Mohammadi, M., Gorde, K., Jajodia, S., Park, H., and Kim, Y. (2018).
\newblock Data synthesis based on generative adversarial networks.
\newblock {\em arXiv preprint arXiv:1806.03384}.

\bibitem[Park and Ghosh, 2014]{park2014pegs}
Park, Y. and Ghosh, J. (2014).
\newblock Pegs: Perturbed gibbs samplers that generate privacy-compliant synthetic data.
\newblock {\em Trans. Data Priv.}, 7(3):253--282.

\bibitem[Popov et~al., 2019]{popov2019neural}
Popov, S., Morozov, S., and Babenko, A. (2019).
\newblock Neural oblivious decision ensembles for deep learning on tabular data.
\newblock In {\em International Conference on Learning Representations}.

\bibitem[Rabiner, 1989]{rabiner1989tutorial}
Rabiner, L.~R. (1989).
\newblock A tutorial on hidden markov models and selected applications in speech recognition.
\newblock {\em Proceedings of the IEEE}, 77(2):257--286.

\bibitem[Rigaki and Garcia, 2020]{rigaki2020survey}
Rigaki, M. and Garcia, S. (2020).
\newblock A survey of privacy attacks in machine learning.
\newblock {\em arXiv preprint arXiv:2007.07646}.

\bibitem[Rosenblatt et~al., 2020]{rosenblatt2020differentially}
Rosenblatt, L., Liu, X., Pouyanfar, S., de~Leon, E., Desai, A., and Allen, J. (2020).
\newblock Differentially private synthetic data: Applied evaluations and enhancements.
\newblock {\em arXiv preprint arXiv:2011.05537}.

\bibitem[Salimans and Kingma, 2016]{salimans2016weight}
Salimans, T. and Kingma, D.~P. (2016).
\newblock Weight normalization: A simple reparameterization to accelerate training of deep neural networks.
\newblock {\em Advances in neural information processing systems}, 29.

\bibitem[Shokri and Shmatikov, 2015]{shokri2015privacy}
Shokri, R. and Shmatikov, V. (2015).
\newblock Privacy-preserving deep learning.
\newblock In {\em Proceedings of the 22nd ACM SIGSAC conference on computer and communications security}, pages 1310--1321.

\bibitem[Shokri et~al., 2017]{shokri2017membership}
Shokri, R., Stronati, M., Song, C., and Shmatikov, V. (2017).
\newblock Membership inference attacks against machine learning models.
\newblock In {\em 2017 IEEE symposium on security and privacy (SP)}, pages 3--18. IEEE.

\bibitem[Song and Mittal, 2021]{song2021systematic}
Song, L. and Mittal, P. (2021).
\newblock Systematic evaluation of privacy risks of machine learning models.
\newblock In {\em 30th USENIX Security Symposium (USENIX Security 21)}, pages 2615--2632.

\bibitem[Srivastava et~al., 2014]{srivastava2014dropout}
Srivastava, N., Hinton, G., Krizhevsky, A., Sutskever, I., and Salakhutdinov, R. (2014).
\newblock Dropout: a simple way to prevent neural networks from overfitting.
\newblock {\em The journal of machine learning research}, 15(1):1929--1958.

\bibitem[Tantipongpipat et~al., 2021]{tantipongpipat2021differentially}
Tantipongpipat, U.~T., Waites, C., Boob, D., Siva, A.~A., and Cummings, R. (2021).
\newblock Differentially private synthetic mixed-type data generation for unsupervised learning.
\newblock In {\em 2021 12th International Conference on Information, Intelligence, Systems \& Applications (IISA)}, pages 1--9. IEEE.

\bibitem[Tian et~al., 2021]{tian2021evolutionary}
Tian, Y., Si, L., Zhang, X., Cheng, R., He, C., Tan, K.~C., and Jin, Y. (2021).
\newblock Evolutionary large-scale multi-objective optimization: A survey.
\newblock {\em ACM Computing Surveys (CSUR)}, 54(8):1--34.

\bibitem[Torfi et~al., 2022]{torfi2022differentially}
Torfi, A., Fox, E.~A., and Reddy, C.~K. (2022).
\newblock Differentially private synthetic medical data generation using convolutional gans.
\newblock {\em Information Sciences}, 586:485--500.

\bibitem[Tram{\`e}r et~al., 2016]{tramer2016stealing}
Tram{\`e}r, F., Zhang, F., Juels, A., Reiter, M.~K., and Ristenpart, T. (2016).
\newblock Stealing machine learning models via prediction $\{$APIs$\}$.
\newblock In {\em 25th USENIX security symposium (USENIX Security 16)}, pages 601--618.

\bibitem[Tseng and Wu, 2020]{tseng2020compressive}
Tseng, B.-W. and Wu, P.-Y. (2020).
\newblock Compressive privacy generative adversarial network.
\newblock {\em IEEE Transactions on Information Forensics and Security}, 15:2499--2513.

\bibitem[Verma et~al., 2021]{verma2021comprehensive}
Verma, S., Pant, M., and Snasel, V. (2021).
\newblock A comprehensive review on nsga-ii for multi-objective combinatorial optimization problems.
\newblock {\em IEEE access}, 9:57757--57791.

\bibitem[Wang et~al., 2021]{wang2021variational}
Wang, K.-C., Fu, Y., Li, K., Khisti, A., Zemel, R., and Makhzani, A. (2021).
\newblock Variational model inversion attacks.
\newblock {\em Advances in Neural Information Processing Systems}, 34:9706--9719.

\bibitem[Webster et~al., 2019]{webster2019detecting}
Webster, R., Rabin, J., Simon, L., and Jurie, F. (2019).
\newblock Detecting overfitting of deep generative networks via latent recovery.
\newblock In {\em Proceedings of the IEEE/CVF Conference on Computer Vision and Pattern Recognition}, pages 11273--11282.

\bibitem[Wierzbicki, 1980]{wierzbicki1980use}
Wierzbicki, A.~P. (1980).
\newblock The use of reference objectives in multiobjective optimization.
\newblock In {\em Multiple criteria decision making theory and application}, pages 468--486. Springer.

\bibitem[Xie et~al., 2018]{xie2018differentially}
Xie, L., Lin, K., Wang, S., Wang, F., and Zhou, J. (2018).
\newblock Differentially private generative adversarial network.
\newblock {\em arXiv preprint arXiv:1802.06739}.

\bibitem[Xu et~al., 2019]{xu2019modeling}
Xu, L., Skoularidou, M., Cuesta-Infante, A., and Veeramachaneni, K. (2019).
\newblock Modeling tabular data using conditional gan.
\newblock {\em Advances in Neural Information Processing Systems}, 32.

\bibitem[Yang et~al., 2019]{yang2019neural}
Yang, Z., Zhang, J., Chang, E.-C., and Liang, Z. (2019).
\newblock Neural network inversion in adversarial setting via background knowledge alignment.
\newblock In {\em Proceedings of the 2019 ACM SIGSAC Conference on Computer and Communications Security}, pages 225--240.

\bibitem[Yeh and Lien, 2016]{data_taiwan}
Yeh, I.-C. and Lien, C.-H. (2016).
\newblock default of credit card clients data set.
\newblock \url{https://archive.ics.uci.edu/ml/datasets/default+of+credit+card+clients}.

\bibitem[Yeom et~al., 2017]{yeom2017unintended}
Yeom, S., Fredrikson, M., and Jha, S. (2017).
\newblock The unintended consequences of overfitting: Training data inference attacks.
\newblock {\em arXiv preprint arXiv:1709.01604}, 12.

\bibitem[Yoon et~al., 2019]{yoon2019time}
Yoon, J., Jarrett, D., and Van~der Schaar, M. (2019).
\newblock Time-series generative adversarial networks.
\newblock {\em Advances in neural information processing systems}, 32.

\bibitem[Zhang et~al., 2017]{zhang2017privbayes}
Zhang, J., Cormode, G., Procopiuc, C.~M., Srivastava, D., and Xiao, X. (2017).
\newblock Privbayes: Private data release via bayesian networks.
\newblock {\em ACM Transactions on Database Systems (TODS)}, 42(4):1--41.

\bibitem[Zhang and Li, 2007]{zhang2007moea}
Zhang, Q. and Li, H. (2007).
\newblock Moea/d: A multiobjective evolutionary algorithm based on decomposition.
\newblock {\em IEEE Transactions on evolutionary computation}, 11(6):712--731.

\bibitem[Zhang et~al., 1996]{zhang1996birch}
Zhang, T., Ramakrishnan, R., and Livny, M. (1996).
\newblock Birch: an efficient data clustering method for very large databases.
\newblock {\em ACM sigmod record}, 25(2):103--114.

\bibitem[Zhang et~al., 2020]{zhang2020secret}
Zhang, Y., Jia, R., Pei, H., Wang, W., Li, B., and Song, D. (2020).
\newblock The secret revealer: Generative model-inversion attacks against deep neural networks.
\newblock In {\em Proceedings of the IEEE/CVF Conference on Computer Vision and Pattern Recognition}, pages 253--261.

\bibitem[Zhang et~al., 2022]{zhang2022membership}
Zhang, Z., Yan, C., and Malin, B.~A. (2022).
\newblock Membership inference attacks against synthetic health data.
\newblock {\em Journal of biomedical informatics}, 125:103977.

\bibitem[Zhao et~al., 2021]{zhao2021ctab}
Zhao, Z., Kunar, A., Birke, R., and Chen, L.~Y. (2021).
\newblock Ctab-gan: Effective table data synthesizing.
\newblock In {\em Asian Conference on Machine Learning}, pages 97--112. PMLR.

\bibitem[Zhou et~al., 2011]{zhou2011multiobjective}
Zhou, A., Qu, B.-Y., Li, H., Zhao, S.-Z., Suganthan, P.~N., and Zhang, Q. (2011).
\newblock Multiobjective evolutionary algorithms: A survey of the state of the art.
\newblock {\em Swarm and evolutionary computation}, 1(1):32--49.

\bibitem[Zhu et~al., 2017]{zhu2017unpaired}
Zhu, J.-Y., Park, T., Isola, P., and Efros, A.~A. (2017).
\newblock Unpaired image-to-image translation using cycle-consistent adversarial networks.
\newblock In {\em Proceedings of the IEEE international conference on computer vision}, pages 2223--2232.

\bibitem[Zitzler and Thiele, 1999]{zitzler1999multiobjective}
Zitzler, E. and Thiele, L. (1999).
\newblock Multiobjective evolutionary algorithms: a comparative case study and the strength pareto approach.
\newblock {\em IEEE transactions on Evolutionary Computation}, 3(4):257--271.

\end{thebibliography}

% \appendix

% \input{Sections/secz_a}
% \input{Sections/secz_b}

\end{document}